# Electric field-dependent conductivity as probe for charge carrier delocalization and morphology in organic semiconductors

Morteza Shokrani, Felix Maximilian Graf, Anton Kompatscher, Dennis Derewjanko, and Martijn Kemerink*

Institute for Molecular Systems Engineering and Advanced Materials (IMSEAM), Heidelberg University, 69120 Heidelberg, Germany

*Corresponding author. email: martijn.kemerink@uni-heidelberg.de

**Abstract**

The charge carrier localization length $\alpha$ is a crucial, yet often ignored parameter of conjugated polymers that exponentially influences electronic conductivity. Here, we argue it is a unique proxy of the energy landscape as determined by sample morphology and experienced by mobile charges. To determine $\alpha$, we use that in disordered organic semiconductors, slow thermalization of charge carriers after excitation, e.g. by hopping in a finite electric field, can lead to an effective electronic temperature ($T_{eff}$) exceeding the lattice temperature, thereby enhancing conductivity. We experimentally probe this effect by combining temperature- and field-dependent conductivity measurements for a range of representative conjugated polymers, using different dopants, doping protocols and doping concentrations. We find that in the high-field regime ($F > 10^6$V/m), $T_{eff}$ exhibits distinct trends vs. structural order and doping level, which can be used to extract (effective) localization lengths ranging from ~1nm in fully amorphous systems to over ~10nm in highly ordered polymers. Tight-binding and kinetic Monte Carlo simulations are used to connect measured values to morphological properties and to rule out alternative explanations. Our results demonstrate that finite-field conductivity measurements provide a powerful probe of a characteristic length scale of charge transport that is complementary to conventional structural characterization.

**Introduction**

Organic semiconductors (OSC are vital for next-generations of low-cost, flexible electronic devices such as solar cells (OPVs),[1] light-emitting diodes (OLEDs),[2] sensors,[3] field-effect transistors (OFETs),[4,5] and neuromorphic devices,[6] where efficient charge transport is crucial for performance. In typical organic materials, charge transport is governed by hopping, a thermally activated, incoherent process where carriers tunnel between localized electronic states residing on individual or small numbers of molecules/repeat units.[7,8] This strong Anderson-type localization ultimately arises from the weak van der Waals interactions between molecules, which leads to disorder that prevents the formation of continuous electronic bands seen in traditional crystalline inorganic semiconductors.[9] Further factors contributing to disorder include charged and neutral impurities, including dopants, structural defects, chemical imperfections, and in general the random spatial arrangement of the constituent molecules, making disorder an intrinsic characteristic of most OSC[10–12].

Although many aspects of charge transport in disordered OSC have been investigated in great experimental and theoretical detail, this does not hold for the static charge carrier localization length $\alpha$ that is one of the key parameters characterizing the electronic wavefunctions. We note that dynamic (de)localization has recently received significant attention in the context of the transient localization framework that is applicable to highly ordered OSC and especially single crystals, which are of smaller practical relevance and not the topic of this work.[13,14] While it is appreciated that, for instance, charge separation in OPVs[15–17] and conductivity in both doped and intrinsic OSC[9,14,18,19] depend critically on the spatial extend of the electronic wavefunctions, direct and systematic measurements of the associated length scale are rare.[18,20] An experimentally validated quantitative or even qualitative understanding of how morphology, doping, and molecular ordering influence the localization length seems to be lacking. This is especially urgent – and surprising – as hopping rates, and hence mobility and conductivity, depend strongly, typically exponentially, on the localization length.[7,8]

A further critical implication of the inherent disorder in typical OSC is the slow thermalization of photo- or electrically-excited charge carriers.[7,21,22] Unlike the rapid thermalization in inorganic materials, the rugged energy landscape created by disorder causes excited, "hot", carriers to retain any excess energy in the global density of states (DOS) for longer periods.[7,23] This leads to field-enhanced conductivities in devices such as OLEDs, OPVs and OFETs in which high electric fields around $10^6$ – $10^7$ V/m are routinely encountered under operating conditions. While of evident importance in its own right, the field dependence of electronic conductivity also offers an avenue to probe the (effective) localization length as will be discussed herein.

Here, we systematically study field-dependent transport in a series of organic semiconductors with varying degrees of aggregation and amorphous character. Our results reveal a strong dependence of the extracted static localization length on the degree of crystallinity, doping method, and dopant type. Furthermore, we investigate the field dependence of the localization length using a tight-binding model, which predicts a slight reduction in $\alpha$ with increasing field. This subtle effect may account for minor deviations from the theoretically expected trends at moderate to high fields. Overall, our findings demonstrate that field-dependent conductivity measurements and the effective temperature framework provide valuable tools for probing the length scales that dominate the electronic transport in disordered OSC.

To put this work in perspective, we first provide a brief overview of previous works on, and concepts used in the interpretation of, field-dependent conductivity in OSC. Early studies in disordered OSC have reported a Pool-Frenkel-type field dependence of conductivity described by the relation $\sigma \propto \exp(\sqrt{F})$, observed at electric fields ranging from $10^6$ to $10^8$ V/m.[24,25] In this model, charge carriers are assumed to be coulombically trapped in the material. A large electric field effectively lowers the potential barrier that electrons must overcome to escape these trap states and move into the conduction band. In contrast, at very high fields, phenomena such as Wannier-Stark[26] localization may emerge, which could reduce charge carrier mobility by further localizing states, which actually runs counter to the typically observed increased conductivity with field, as described by e.g. the Poole-Frenkel effect. Furthermore, there exists a significant body of works using variants of the Gaussian disorder model[27] to study $\sigma(F)$, typically finding nonlinear increases at finite fields.[27–31] Unfortunately, due to their high computational demands and the significant number of adjustable parameters, these approaches are less suited for the interpretation of experimental data.

An alternative framework for understanding charge transport under high electric fields in disordered systems was suggested by Marianer-Shklovski.[21] In this model, that was based on numerical simulations, it is argued that the combination of high electric fields $F$ and slow thermalization can drive the system out of equilibrium.[23] The resulting non-equilibrium energy distribution can heuristically be described by an effective electronic temperature,

$$T_{eff} = \left( T_{latt}^\beta + \left( \gamma \frac{q\alpha F}{k_B} \right)^\beta \right)^{\frac{1}{\beta}}, \qquad (1)$$

where $q$ is the elementary charge, $\alpha$ is the charge localization length, representing the decay rate of the wavefunction of a tunneling charge carrier, $k_B$ is the Boltzmann constant, and $\beta$ and $\gamma$ are phenomenological constants. The consequence of Eq. 1 is that the field dependent conductivity $\sigma(T_{latt}, F)$ could be expressed in terms of the low-field conductivity as $\sigma(T_{eff}(T_{latt}, F))$.

While the Marianer–Shklovskii model captures key features of field-enhanced transport, it lacks a clear physical interpretation of the effective temperature. In a recent publication, we introduced a more physically grounded formulation of the effective temperature based on a balance between energy gain by motion along the electric field and energy loss by hopping to energetically lower states.[32] This heat balance (HB) model leads to the following expression for the effective temperature, cf. SI Section 3:

$$T_{eff} = \frac{T_{latt} + \left( T_{latt}^2 + \left( \gamma \frac{q\alpha F}{k_B} \right)^2 \right)^{0.5}}{2} \qquad (2)$$

For the commonly used value $\beta = 2$, Eqs. 1 and 2 are functionally indistinguishable within typical experimental error. In both expressions, the localization length is the only free parameter, making a measurement of $T_{eff}(F)$ potentially to a direct probe of $\alpha$ in disordered systems. It is, in this context, important that we could experimentally demonstrate that the effective temperature is not just a mathematical construct that allows to heuristically map a field dependence on a temperature dependence but reflects the true characteristic temperature of the electron (or hole) distribution of occupied states.[33]

By utilizing the effective temperature concept, we are therefore able to extract localization lengths from electrical measurements and relate them systematically to material properties such as crystallinity, aggregation, and doping. This provides a valuable bridge between microscopic structure and macroscopic charge transport.

**Results and Discussion**

To investigate the electric field dependent conductivity in OSC, we selected a set of material systems with well-established and structurally distinct characteristics. These systems differ in terms of degree of crystallinity and are doped with different dopants and doping methods, factors which are known to strongly influence the electronic properties of thin films. Hence, our goal is not to re-establish well-known trends in morphology, but rather to exploit this knowledge to investigate their direct impact on measured localization lengths and, in general, on field-driven charge transport under device-relevant electric fields. Our material set includes regioregular (rr) and regiorandom (rra) P3HT to assess the influence of backbone ordering, bulk and sequentially doped P3HT to isolate the effects of the doping method, PBTTT, P3HT and Super Yellow (SY) to examine the role of liquid- and semi-crystalline and amorphous morphologies, and P3HT doped with either $F_4$TCNQ or Magic Blue (MB) to probe dopant–host interactions in both ordered and disordered morphologies. Structures and systematic names are provided in Figure S1 and Table S1 of the Supporting Information (SI).

To apply the effective temperature framework, it is necessary to first establish a reference temperature dependence of the (ohmic) conductivity under low-field conditions. This baseline allows us to map the high-field conductivity onto an equivalent temperature via the relation

$$\sigma(T,F) = \sigma(T_{eff}(T,F), 0), \qquad (3)$$

as discussed above. Figure 1a presents the measured temperature-dependent ohmic conductivity for the selected material systems. All materials exhibit thermally activated behavior, where conductivity increases with temperature, which is consistent with hopping transport. The data are well-described by an empirical stretched exponential form

$$\sigma = \sigma_0 \exp\left(-\left(\frac{T_0}{T}\right)^\delta\right), \qquad (4)$$

where $\sigma_0$ and $T_0$ are fitting parameters and $\delta$ is a stretching exponent that reflects the dominant hopping mechanism. To extract $\delta$, we employ the Zabrodskii analysis, in which the derivative W=$d \log \sigma / d \log T$ is plotted as $\log W$ versus $\log T$ (Figure 1b). This method yields linear trends whose slopes correspond to the exponent $\delta$, providing a reliable and widely used technique for identifying charge transport regimes in disordered materials.

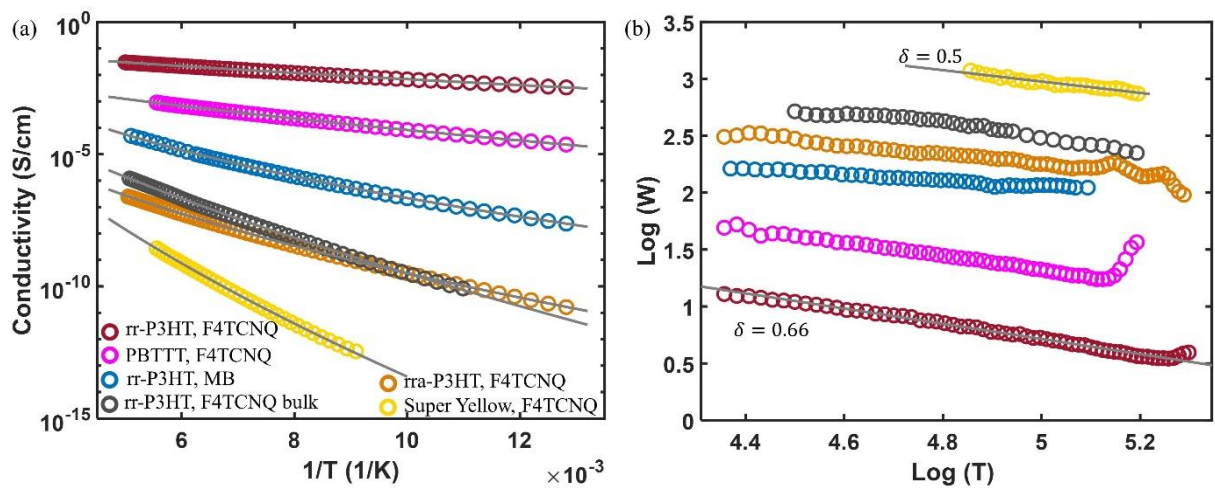

**Figure 1.** (a) Low-field electrical conductivity as a function of $1/T$ in the temperature range 78 K – 200 K. Solid lines are fits to Eq. 4 using the exponents determined in panel b. (b) Zabrodskii plots of the same data.

Detailed fitting parameters from Eq. 4 are summarized in Table 1. While some of the extracted temperature exponents align with well-established transport mechanisms, such as $\delta = 0.25$ for 3D (Mott) variable range hopping (VRH) in a constant DOS and $\delta = 0.5$ for Efros-Shklovskii VRH in presence of a soft Coulomb gap,[34] many values fall within an intermediate range of $0.25 \leq \delta \leq 0.66$. This heterogeneity suggests a broader range of charge transport behaviors than conventional VRH models alone can explain – which is to be expected as the mentioned models typically assume homogeneous media and only hold in specific parts of parameter space.

Although a full discussion of the stretching exponents is beyond the scope of the present work, we do want to highlight a particular trend for sequentially $F_4TCNQ$-doped rr-P3HT that shows a notable increase in $\delta$ from 0.35 to 0.66 with increasing doping levels that cannot simply be explained by the opening of a Coulomb gap. For this material system, it is known that the dopant gets preferentially incorporated in the crystalline phases of the P3HT thin film, leading to a heterogeneous material that consists of highly doped crystalline regions separated by largely undoped amorphous phases.[35] Although this class of models has not been widely applied to organics, this situation is not unlike what is considered in the Ping Sheng model for granular metals, in which hopping occurs between conducting regions embedded within an insulating matrix.[36–38] In this framework, the value of $\delta$ depends on the size of the conductive domains (e.g., doped P3HT aggregates) and their spatial separation within the insulating host (e.g., amorphous P3HT) and increases for increasing mean size and decreasing separation.[39] Hence, in our system, the increase in $\delta$ with doping could possibly indicate the growth of conductive aggregates due to progressing dopant infiltration.[40–42] However, this interpretation is speculative and would require direct structural characterization for validation, which lies beyond the scope of the current study.

**Table 1**. Parameters determined from low-field temperature-dependent conductivity measurements for all investigated materials. For some cases, doping concentrations had to be de- (e.g. PBTTT) or increased (Super Yellow) in order to enable reliable measurements. This choice was dictated by practical constraints: materials with intrinsically high mobilities can reach very high conductivities at moderate carrier densities, which in turn causes excessive Joule heating under applied fields. In contrast, in materials with low mobilities, a too low doping level would yield conductivities below our detection limit, particularly at low temperatures. The doping concentrations were therefore optimized individually for each system to balance these effects.

|  | dopant | concentration | $\delta$ | $T_0$ | $\sigma_0$ |
|---|---|---|---|---|---|
| P3HT RR | $F_4$TCNQ/ sequential | 2 mg/ml | 0.66 | 1168 | 0.53 |
| P3HT RR | $F_4$TCNQ/ sequential | 0.8 mg/ml | 0.5 | 3939 | 0.8 |
| P3HT RR | $F_4$TCNQ/ sequential | 0.2 mg/ml | 0.35 | $2.49 \times 10^5$ | 51.8 |
| P3HT RRa | $F_4$TCNQ/ sequential | 2 mg/ml | 0.4 | $4.16 \times 10^5$ | 464 |
| P3HT RR | $F_4$TCNQ/ bulk | 100 to 1 | 0.6 | $2.21 \times 10^4$ | 30.5 |
| P3HT RR | MB/ sequential | 0.2 mg/ml | 0.25 | $1.52 \times 10^8$ | $3.78 \times 10^8$ |
| PBTTT | $F_4$TCNQ/ sequential | 0.023 mg/ml | 0.6 | 3166 | 0.2343 |
| Super yellow | $F_4$TCNQ/ sequential | 5 mg/ml | 0.5 | $1.94 \times 10^5$ | $5.04 \times 10^5$ |

After establishing the temperature dependence of the conductivity, we measure the electric field dependent conductivity at constant lattice temperature using either pulsed or fast direct current (DC) measurements to avoid possible Joule heating at high electric fields, see Figure S2 and the corresponding discussion in the SI for details. Figure 2a shows the field dependence of the conductivity for three selected systems of PBTTT (liquid crystalline), P3HT (semi-crystalline) and SY (amorphous), all sequentially doped with $F_4$TCNQ; the full data for all systems measured is shown in Figure S4 of the SI. At the lowest fields, all three systems exhibit ohmic behavior, characterized by a field-independent conductivity. As the field increases, a nonlinear rise in conductivity is observed, which is an indication of field-activated transport. Notably, SY exhibits the lowest conductivity despite its higher doping level, attributable to its highly disordered amorphous morphology, which suppresses charge mobility. The negative differential conductivity of this material at low intermediate fields is attributed to so-called returns, which are parts of the percolating network running against the field and are expected for high-disorder systems.[43,44] We note that PBTTT shows a lower conductivity than P3HT, which is attributable to the lower doping level that was needed for the experiments, as explained in the caption to Table 1.

The corresponding effective temperature values, extracted using Eq. 3, are shown in Figure 2b. In the ohmic regime, the effective temperature remains constant and equal to the lattice temperature, as expected. At higher fields, the effective temperature increases monotonically, with distinct trends observed across the three materials. Notably, PBTTT exhibits higher effective temperatures than P3HT at equivalent fields, which may reflect its more ordered morphology that supports extended electronic delocalization. In contrast, SY shows the lowest induced effective temperature, consistent with its amorphous nature and concomitantly highly localized electronic states.

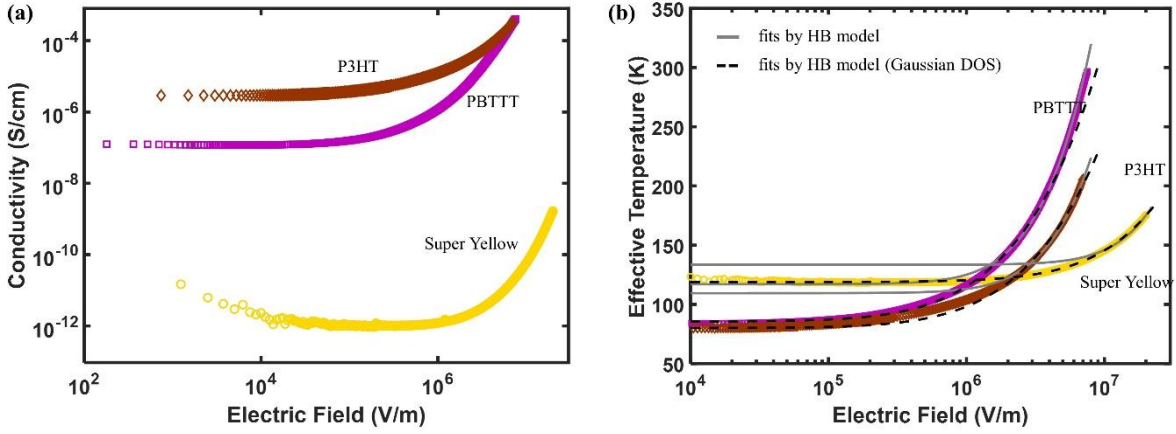

**Figure 2**. (a) Measured electrical conductivity and (b) derived effective temperature as a function of electric field of sequentially doped rr-P3HT (0.2 mg/ml), PBTTT and SY at 80 K. Conductivity fitting parameters are given in Table 1. In b), the grey solid lines are high-field fits to Eq. 2 while black dashed lines are fits to Eq. 5 with $\gamma_0 \alpha F_0^{0.33} =$ 9.86×10⁻⁷, 6.96×10⁻⁷and 2.19×10⁻⁷ for PBTTT, P3HT and SY, respectively. The SY measurement was performed at 120 K since the conductivity at lower temperatures was too low for reliable measurements.

To further analyze the effective temperature results, we use the HB model (Eq. 2) to fit the field dependence of the effective temperature (solid lines in Figure 2b) at intermediate and high electric fields. However, both the HB and the MS models fail to accurately describe the data in the lower field regime, as shown by the solid lines that extrapolate the high-field fits to the low-field regime. In the following, we shall therefore subsequently discuss the low- and high-field regimes.

**Low-field regime**

Similar deviations from the functional form Eq. 1, or, virtually equivalently, Eq. 2, as shown in Figure 2b have been observed before in amorphous carbon nitride and were interpreted in terms of a field-enhanced bandtail hopping model in which the effective temperature concept describes the non-equilibrium occupation probability of tail states in an exponential DOS.[45] At high fields, this model predicts a scaling of conductivity with $F^{0.67}/T$, which can be integrated into Eq. 1 to yield

$$T_{eff} = \frac{T_{latt} + \left(T_{latt}^2 + \left(\gamma_0 \left(\frac{F_0}{F}\right)^{0.33} \frac{q\alpha F}{k_B}\right)^2\right)^{0.5}}{2}. \quad (5)$$

Fits to Eq. 5 are shown in Figure 2b as dashed lines and indeed provide a satisfactory fit over the full field range. However, to validate Eq. 5 for parameters that are typical to doped OSC, we performed numerical kinetic Monte Carlo simulations for charges hopping in either a gaussian or an exponential DOS.[46] We note that the same physics that was explicitly accounted for in the derivation of the scaling factor $F^{0.67}/T$, most notably the field-driven shift of the demarcation energy, are implicitly accounted for, i.e., they are emergent, in the kMC model. The results are shown in Figure 3a and can be accurately described with the expressions Eq. 1 and 2 as they do not show the rise around 10⁵ V/m that is observed in most experiments. Hence, we conclude that any agreement between Eq. 5 and the experimental data must be phenomenological and result from physical effects that are not accounted for in the derivation of Eq. 5. In passing we note that the comparison between Eqs. 1 and 2 in Figure 3a fixes the value of $\gamma$ (to ~1.4) in Eq. 2 that was not exactly defined in the derivation of the heat balance model.

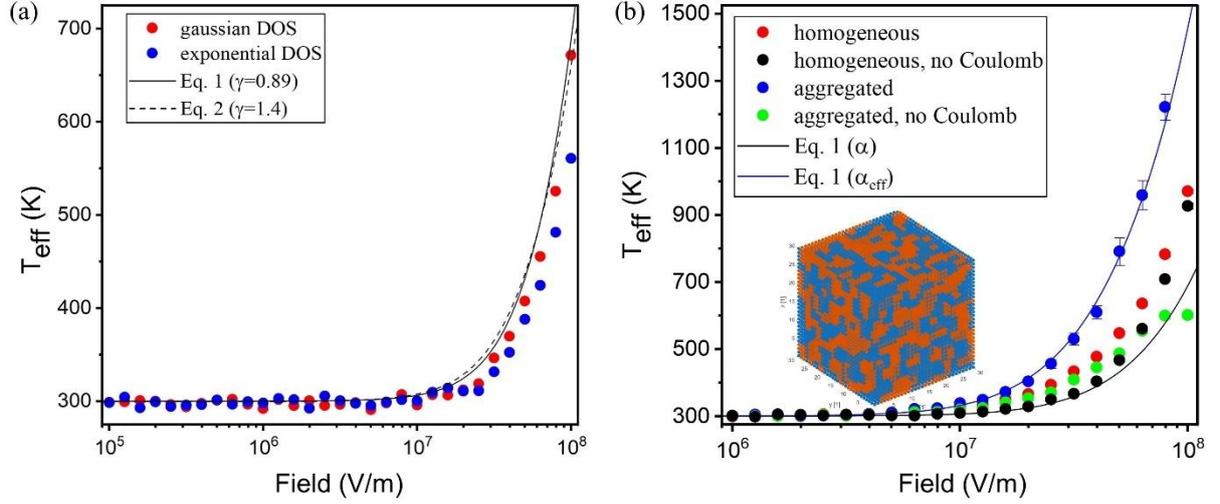

**Figure 3.** Numerical calculation of the effective electronic temperature by kinetic Monte Carlo. (a) Calculations for homogeneous materials with gaussian and exponential DOS on a random lattice, ignoring Coulomb interactions beyond the on-site repulsion. Parameters used are $T_{latt} = 300$ K, $\alpha = 0.6$ nm, energetic disorder 75 meV, relative charge carrier density 1%, $30^3$ sites, periodic boundary conditions and mean inter-site distance 1.8 nm. The lines are calculated from Eq. 1 and Eq. 2 using the kMC input parameters and $\gamma$-values given in the legend. (b) Calculations for Gaussian DOS on regular lattice, using either a homogeneous or aggregated (cf. inset) morphology, w/ or w/o full Coulomb interaction as noted in the legend. Lines are calculated from Eq. 1 with $\gamma = 0.89$ and $\alpha_{eff} = 2.3\alpha$. Energetic disorder and attempt-to-hop rate of the crystalline/amorphous phase are 50/75 meV and $3\times10^{11}/3\times10^{10}$ s$^{-1}$, respectively. A 0.3 eV offset between the HOMO energies of the crystalline and amorphous phases is used because the amorphous phase typically exhibits a deeper HOMO level.[35,47,48]. All other parameters as in panel (a).

A possible reason for the unexpectedly strong low-field dependence in the experiments is a field-dependent localization length. Both in the derivation of Eq. 5 and in the kMC simulations, as well as in the considerations that led to Eqs. 1 and 2, the localization length was assumed to be constant. This need not be so, and it is conceivable that $\alpha$ shrinks with increasing field due to, e.g., Stark localization. To investigate the influence of the electric field on the wavefunction localization, we set up a simple tight binding model for an organic semiconductor including the electric field $F$ (in the z-direction) as follows:

$$H = \sum_i (\epsilon_i + \langle i|e\boldsymbol{F}\boldsymbol{r}_i|i\rangle)a_i^\dagger a_i - \sum_{\langle i,j\rangle} \hbar v_0 S_{ij} a_i^\dagger a_j + h.c. \quad (6)$$

The on-site energies $\epsilon_i$ are drawn from a gaussian distribution with $\sigma_{DOS} = 75$ meV disorder and zero mean. $\hbar v_0 S_{ij}$ is the transfer integral that is proportional to the overlap $S_{ij}$ between states $|i\rangle, |j\rangle$ assumed to be ellipsoidal s-type orbitals, making the transfer direction- and orientation-dependent. We use a relatively high value of $\hbar v_0 \cong 125$ meV to obtain a reasonable upper limit for the delocalization and hence the field dependence of the wavefunctions. The correction for the electric field shifts the on-site energies according to the position $\mathbf{r}_i$ of the sites $i$. The underlying morphology is based on a model for partially ordered (aggregated) entangled polymer chains on a 20x20x40 simple cubic lattice with lattice constant $a_{NN} = 0.5$ nm; see SI section 5, Figures S5, S6 and Ref. [49] for details on the model.

After setting up the Hamiltonian, the eigenvalue problem $HV = EV$ for the eigenvectors $V$ is solved. The localization length of each eigenstate $V_i$ is extracted via the modified inverse participation ratio (IPR) expression

$$\alpha(E_i) = \left(\sqrt[3]{IPR(E_i)} - 1\right) \cdot a_{NN} + \lambda_s \tag{7}$$

$$IPR(E_i) = \frac{1}{\sum_n |V_{i,n}|^4}, \tag{8}$$

where $\lambda_s = 0.2$ nm is the mean localization parameter of the s-type basis orbitals and $V_{i,n}$ is the contribution of basis state $n$ to eigenvector $i$.

Figure 4 shows the results of the calculated localization length distribution for three different field strengths. As expected, the localization length is energy-dependent, peaking close to the center of the corresponding DOS, shown by histograms.[19] For the specific parameters used, the absolute values reach up to ~4 nm, indicating that the wavefunctions, especially towards the center of the DOS, in which region also the transport energy resides, can delocalize over multiple polymer repeat units, in agreement with more advanced calculations.[9] At the same time, the $\alpha$-values remain small as compared to the several tens, up to over a hundred nm that would be needed to explain the onset of the field dependence around $10^5$ V/m in Figure 2. We also note that the eigenstates remain localized within the structural aggregates, i.e. the localization lengths remain smaller than or equal to typical aggregate sizes that reach up to ~4 nm (see details in SI section 5). While we do observe a decreasing trend of the localization length as expected from Stark localization, the relative magnitude of the mean localization lengths of the distributions only decreases by ~10% at $F = 10^7$ V/m. It is noted that while the mean localization length decreases, the distributions also get wider and higher-energy states can actually get more delocalized. However, the calculated magnitude of the additional (mean) localization due to the applied electric field remains limited in the considered field range. The tight binding calculations therefore do not support an interpretation of the low-field conductivity data as neither the magnitude nor the field dependence of the delocalization comes close to values that correspond to the phenomenological fit with Eq. 5. However, the calculations do show that minor deviations from a strictly constant localization length may be anticipated at the intermediate- and high-field regimes. In this exploratory study, we will not further consider this effect.

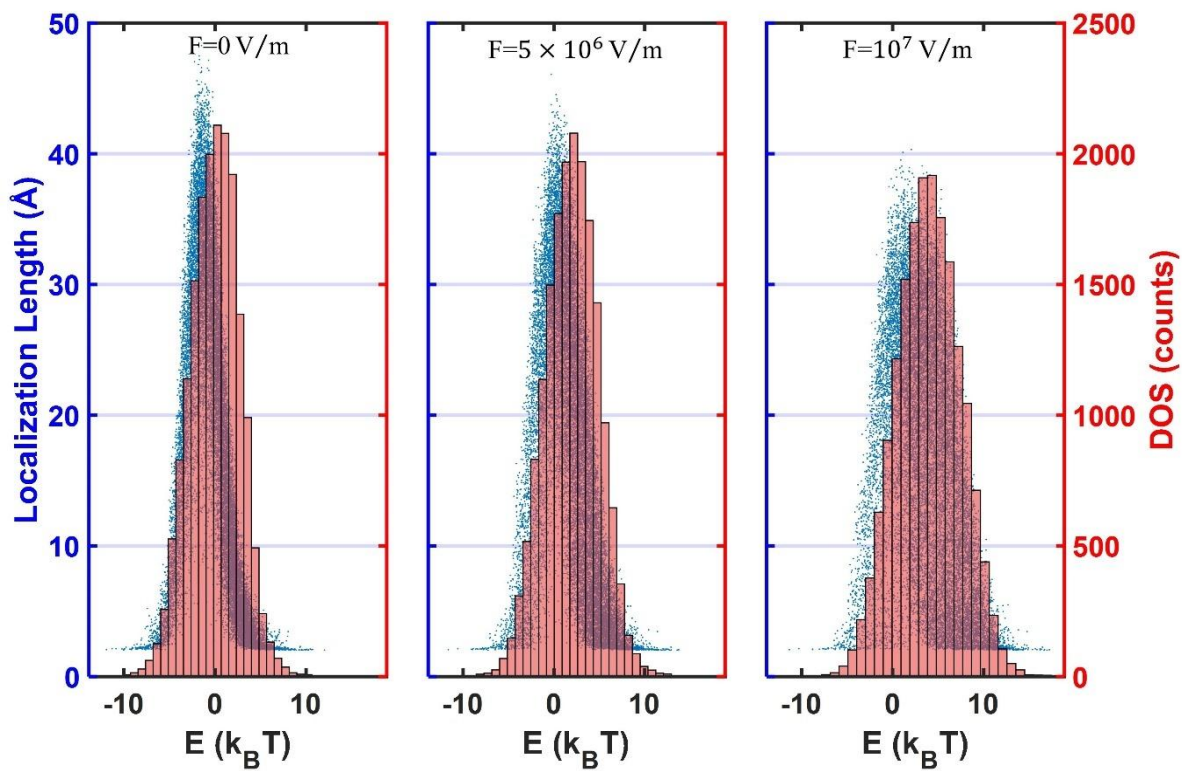

**Figure 4**. Localization length distribution and DOS for different fields calculated from the tight binding model. Blue dots are the values from the IPR, red histograms represent DOS.

A rather different explanation of the surprisingly early onset of a notable field dependence of the conductivity would be a finite and field-dependent contact resistance. Although contact resistances are generally assumed to be small and hence negligible for strongly doped OSC, especially when a high-work function (here: Au) metal is used for hole injection, such effects cannot upfront be ruled out. To examine this scenario, we performed in-operando Kelvin Probe Force Microscopy (KPFM) measurements, the details of which are given in SI Section 2 and Figure S3. These experiments support the presence of moderate contact resistances of up to ~20% of the total device resistance that may affect the field dependence at low (<$10^5$ V/m) fields where conductivity changes are small. At intermediate and high fields, where conductivity changes amount up to several orders of magnitude, these effects are negligible, confirming that our extracted trends reflect intrinsic, bulk material properties. For this reason, we will, in the following, only use the high-field data for further analysis and tentatively attribute the low-field behavior to (field-dependent) contact resistances.

**High-field regime**

The results of the field-dependent conductivity fits to the HB model (Eq. 2) at medium to high electric fields are summarized in Figure 5, presented as bars to allow direct comparison of the extracted localization lengths across different material systems. In view of the explorative nature of this work, the following discussions should be regarded as a first, qualitative investigation of the possibilities of field-dependent conductivity measurements as a tool to study relevant length scales in doped inhomogeneous OSC. An in-depth study of the role of structural properties and processing variables like doping type, method and concentration is beyond the scope of this work and will be topic of further studies.

To explore the role of doping concentration, we sequentially doped rr-P3HT films with $F_4TCNQ$ at concentrations of 0.2, 0.8, and 2 mg/mL, representing (relative) low, medium, and high doping levels, respectively. The extracted localization length increases with increasing doping concentration. Qualitatively, this finding is in full agreement with our previous work Ref. [19] and Figure 4 above: with increasing doping level, the Fermi level shifts upwards, to more delocalized states, while the transport energy stays almost constant. As a consequence, the characteristic, i.e., rate-limiting hops that take place between the Fermi energy and the transport energy involve increasingly delocalized states. However, the absolute values of up to ~12 nm are significantly larger than the (upper limit for the) value ~4 nm that was calculated here and elsewhere.[9,50] Hence, we interpret this number as an 'effective' localization length that reflects the semicrystalline (or later the liquid-crystalline) morphology of P3HT (or later PBTTT), where long-range transport involves both, supposedly fast, intra-aggregate hopping and, supposedly slow, inter-aggregate hopping across the less-conductive amorphous regions separating the aggregates.

Since the tight binding simulations show the absence of tail states (where the Fermi level sits) with localization lengths of the size of a typical aggregate, it seems unlikely that the long-range transport is solely limited by inter-aggregate hops across amorphous barriers. Hence, a scenario in which the hopping charge carrier only accumulates 'tunneling action' in the barriers between the aggregates does not apply here. Previously, such a model was applied to quantum dot arrays and leads to an effective localization length $\alpha_{eff} = ((R+d)/d)\alpha$ with $R$ and $d$ the typical aggregate radius and barrier thickness, respectively.[51] We propose an alternative scenario in which charges pile up at bottlenecks, causing the field to predominantly drop over these barriers, leading to an enhanced field over these barriers. A simple geometrical consideration for a system consisting of amorphous and crystalline regions with typical sizes $L_{am}$ and $L_{crys}$ leads to an enhancement factor of the order $(L_{am} + L_{crys})/L_{am}$, where we have associated the amorphous regions with the bottlenecks. We note that experimentally testing this hypothesis directly is nearly impossible as it requires visualizing barriers, including the absence or presence of tie chains. Hence, we again turn to kMC simulations to test the conceptual viability of this hypothesis. Specifically, we used the cellular automaton that was described in detail in Ref. [35] to generate phase-separated morphologies consisting of crystalline and amorphous material, as illustrated in the inset to Figure 3b. To be able to account for the full Coulomb interaction, simulations had to be run on a regular lattice.

Comparing the blue (aggregated) and red (homogeneous) curves in Figure 3b shows that indeed the presence of a larger characteristic length scale in the system has a pronounced, increasing, effect on the effective temperature. The solid lines are calculated with Eq. 1, which in the case of a regular lattice, and especially in presence of spatial inhomogeneity, should be regarded as a lowest order approximation only. Nevertheless, the extracted field-enhancement factor of ~2.3 is fully consistent with the expectation from the expression above for a phase separated 1:1 mixture, here of amorphous and crystalline matter. The simulations for an aggregated morphology with only on-site Coulomb repulsion (green symbols) largely coincide with those for the homogeneous system, showing that it is the combination of (the full) Coulomb interaction and spatial inhomogeneity that causes the large difference between the blue and red curves and not simply the inhomogeneity. This notion is further substantiated by the simulation of a homogeneous system without long-range Coulomb interaction (black symbols) that only slightly deviates from one with (red). In view of these arguments, we attribute the relatively large localization lengths in Fig. 5 as 'effective' numbers, set by the 'true' localization length, which is determined by quantum mechanics, and a geometric enhancement factor, which reflects the sample morphology. Assuming a 'true' localization length of a few nm then suggests that in the medium and highly (sequentially) doped P3HT, which were processed without specific optimization, the field predominantly falls over relatively narrow regions that are a factor ~5-10 thinner than a typical aggregate size, which seems plausible. We reiterate that a direct experimental test of this interpretation is beyond the capabilities of typical structural characterization tools and hence beyond this work.

Having, tentatively, identified the measured length scale as an effective localization length allows an alternative, or complementary, interpretation of the trend observed in the series low-medium-high doped rr-P3HT. $F_4$TCNQ is known to intercalate within the crystalline phase of P3HT, primarily doping the ordered regions, partly because the amorphous phase typically exhibits a deeper HOMO level.[47,48] This intercalation often induces substantial structural reorganization, potentially increasing crystallite size or even causing the formation of new crystallites.[52] Hence, the larger $\alpha_{eff}$-values for stronger doped rr-P3HT may not only reflect an increase in $\alpha$ due to the Fermi level moving upward, but also an increase in field-enhancement factor due to a growth of the crystalline domains and/or a narrowing of the amorphous barriers. In passing, we note that an interpretation in terms of growing crystalline domains would be consistent with the trend in stretching exponent of the thermal activation, cf. Eq. 4, that was discussed in the context of Figure 1.

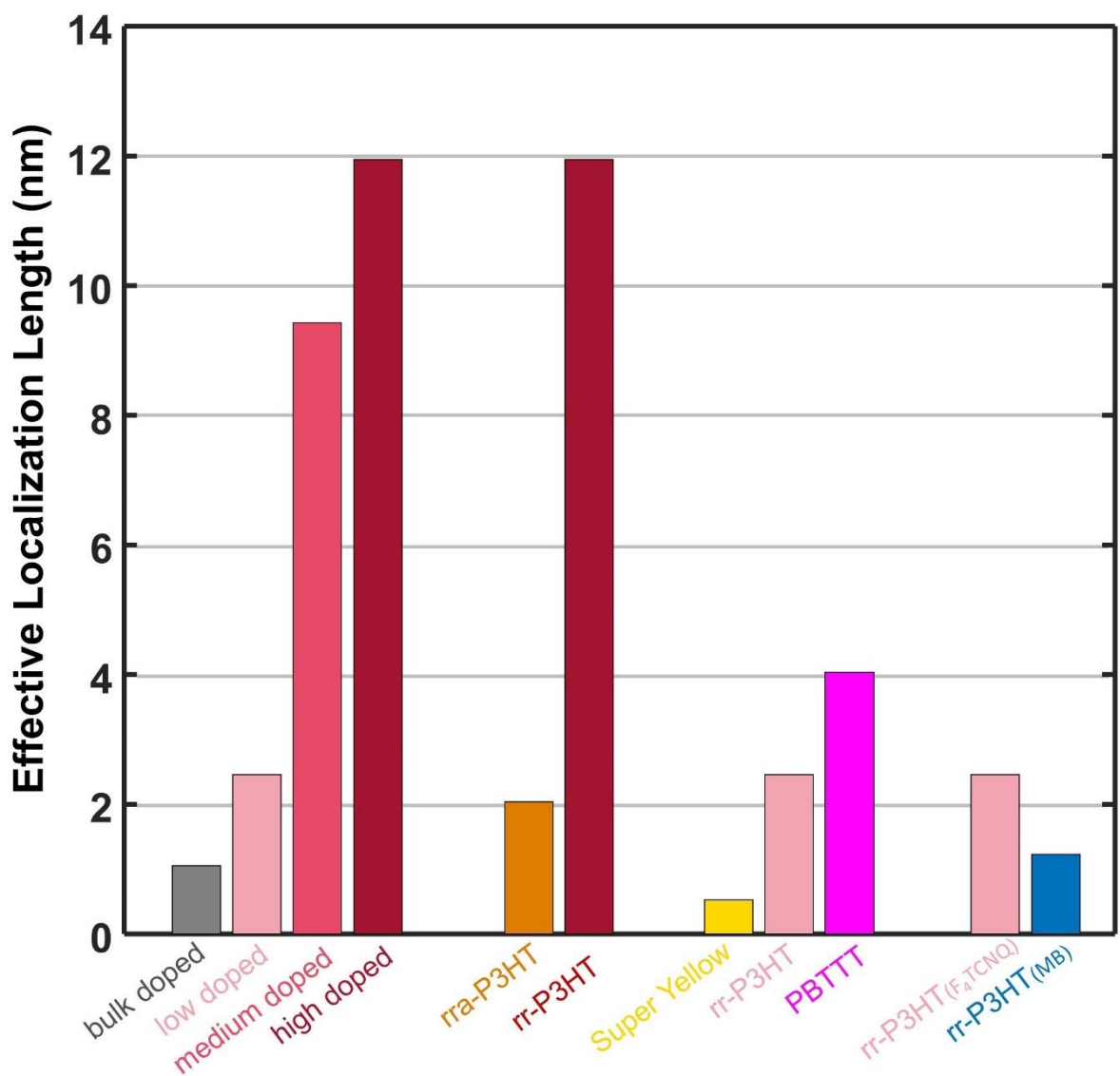

**Figure 5**. Comparison of (effective) localization lengths in different systems. When no semiconductor, dopant and/or doping method are specified, the data are for rr-P3HT, $F_4TCNQ$ and sequential doping, respectively. The comparisons of SY/rr-P3HT/PBTTT and $F_4TCNQ$/MB are for low-doped samples (0.2 mg/ml).

Bearing the previous discussions in mind, several further observations can be made in Figure 5 that, at least qualitatively, corroborate the interpretation of the strength (and onset) of the field dependence of the conductivity in terms of an effective localization length. For instance, it is well-known that the method of doping significantly affects film morphology and thus electronic transport. The data discussed so far were obtained for sequentially doped rr-P3HT, meaning that the pristine polymer is first spin-coated to form a high-quality film that is subsequently doped by spin-coating the dopant from an orthogonal solvent. This allows dopant infiltration without disrupting the polymer matrix. In contrast, bulk doping involves co-dissolving polymer and dopant in a common solvent from which the film is deposited. This method is known to lead to poorer film quality due to the precipitation of highly doped agglomerates in the solution.[53–55] In the present experiments, this leads to bulk-doped films showing significantly lower localization lengths, around 2 nm, that are consistent with values found previously, using temperature-dependent space charge limited conductivity measurements, for intrinsic amorphous polymers and photovoltaic blends.[30]

A similar argument as for the sequential vs. bulk doping process can be made for the marked difference between rra-P3HT and rr-P3HT. In case of rr-P3HT, the regularity of side chain attachment on the P3HT backbone promotes π–π stacking and crystalline ordering. In contrast, the irregular side chain attachment in rra-P3HT results in more amorphous films.[56] Although the different morphologies may, despite the identical doping protocol, lead to different levels of dopant uptake, it seems unlikely that the ~5 times higher localization length in rr-P3HT than in its rra-counterpart solely reflects a concentration difference.

As explained above, the high (low) mobility of PBTTT (SY) necessitated a decrease (increase) in dopant concentration to keep conductivity values within workable ranges. Nevertheless, there is a clear trend in effective localization length for the series SY – rr-P3HT – PBTTT in Figure 5, that would most likely be more pronounced, had the same dopant concentrations been used, cf. the rr-P3HT series. Although PBTTT and P3HT both tend to form morphologies with mixed amorphous and crystallite phases, PBTTT is generally believed to be more prone to form ordered morphologies due to its more rigid backbone, which allows for a better π–π stacking of the polymers.[57,58] On the other hand, Super Yellow (also known as non-red-shifting poly(phenylene-vinylene), NRS-PPV) was originally synthesized for being entirely amorphous.[59] The extracted effective localization lengths follow the expected trend, PBTTT > P3HT > SY, with the value for SY, $\alpha_{eff} \approx 0.8$ nm, likely reflecting the actual localization length $\alpha$, being indicative of fully localized transport in a highly disordered matrix.

Interestingly, the extracted effective localization length for MB-doped rr-P3HT is shorter than for $F_4TCNQ$. Bearing in mind the uncertainties addressed previously, we tentatively interpret this as an indication that MB, unlike $F_4TCNQ$, does not cause any significant structural reorganization, and as such does not lead to an increasing crystallite size. This is in line with the results of previous studies, showing that in contrast to $F_4TCNQ$ that selectively intercalates (and dopes) the crystalline phase of P3HT, Magic Blue predominantly resides within the voids of the amorphous phase of P3HT.[47,48,60] Despite its location, MB effectively dopes both the amorphous and crystalline phases. The latter occurs through an interface-mediated process, where charge transfer happens across the interface between the dopant-rich amorphous regions and the crystalline domains due to the ~0.3 eV energy offset between the HOMO levels in the two phases.[35] As a non-intercalating dopant, MB preserves the crystalline microstructure of P3HT, avoiding disruption of the ordered phase, leading to higher mobilities in otherwise optimized morphologies.[35] The smaller stretching exponent in Table 1, 0.25 for MB vs. 0.35 for $F_4TCNQ$, would be consistent with such a scenario.

**Conclusion and outlook**

In this work, we have demonstrated that combining temperature- and electric field-dependent conductivity measurements is a powerful method for probing charge carrier delocalization in doped organic semiconductors. By applying a physically grounded (and kMC-calibrated) heat balance model, we extracted an effective localization length ($\alpha_{eff}$) that is strongly correlated with material morphology, increasing from ~1 nm in amorphous systems to over 10 nm in highly ordered polymers. This effective localization length was found to be highly sensitive to processing conditions, dopant type and doping method, as well as to the degree of the regioregularity. Our interpretation of $\alpha_{eff}$ as an effective parameter, which reflects not just molecular-scale localization but also larger-scale morphological features like crystalline aggregates and amorphous barriers, is supported by numerical simulations.

Finally, we note that the information contained in the effective localization length is complementary to what can be obtained by conventional structural characterization techniques like GIWAXS, while requiring much simpler infrastructure. While conventional structural characterization tools provide information on packing distances, paracrystallinity and, to a lesser degree, aggregate sizes, they provide little direct information on how these numbers work out in the percolating network that eventually determines conductivity. We think that combined field-and temperature dependent conductivity measurements may be used in the targeted optimization of high-performance doped organic semiconductors by identifying the presence and characteristics (height, spacing) of any rate-limiting barriers in the current pathway. However, further modeling and experimental studies will be required to make this fully quantitative and, ideally, predictive.

**Acknowledgements**

We are grateful to Oliver Meixner for contributions during the early stages of this project. This work has been funded by the German Research Foundation (Deutsche Forschungsgemeinschaft, DFG) project 545050087 and under Germany's Excellence Strategy via the Excellence Cluster 3D Matter Made to Order (EXC-2082/1-390761711). M.K. thanks the Carl Zeiss Foundation for financial support.

**Conflict of interest**

The authors declare no conflict of interest.

**Data availability**

The data supporting this article are included in the manuscript and its Supplementary Information.

Supplementary information to

# Electric field-dependent conductivity as probe for charge carrier delocalization and morphology in organic semiconductors

Morteza Shokrani, Felix Maximilian Graf, Anton Kompatscher, Dennis Derewjanko, and Martijn Kemerink*

Institute for Molecular Systems Engineering and Advanced Materials (IMSEAM), Heidelberg University, 69120 Heidelberg, Germany

*Corresponding author. email: martijn.kemerink@uni-heidelberg.de

## Contents

# 1. Experimental details

**Solution preparation**: All polymers and dopants used in this study were weighed and dissolved in their respective solvents inside an inert, oxygen- and moisture-free glovebox environment. The polymer solutions were stirred overnight on a hotplate at 70 °C to ensure complete dissolution. The detailed compositions of the solutions are summarized in Table S1. The structure of the materials is presented in Figure S1 below.

Table S1. Materials and their respective manufacturer and solvents (ODCB is orthodichlorobenzene, THF is tetrahydrofuran, DCM is dichloromethane, AcN is acetonitrile, CF is chloroform).

| material | chemical name | manufacturer | solvent |
|---|---|---|---|
| rr-P3HT | Poly(3-hexylthiophene-2,5-diyl) | Ossila M108 | ODCB |
| rra-P3HT | Poly(3-hexylthiophene-2,5-diyl) | Sigma Aldrich 510823 | ODCB |
| PBTTT | Poly[2,5-bis(3 dodecylthiophen-2-yl)thieno[3,2-*b*]thiophene] | 1material 888491-18-7 | ODCB |
| Super yellow | poly(phenylene vinylene) | Sigma Aldrich 900438 | ODCB |
| F4TCNQ | 2,3,5,6-Tetrafluoro-7,7,8,8-tetracyanoquinodimethane | Ossila M351 | 4:1 THF:DCM |
| Magic Blue | Tris(4-bromophenyl)ammoniumyl hexachloroantimonate | Sigma Aldrich 230227 | 3:1 AcN:CF |

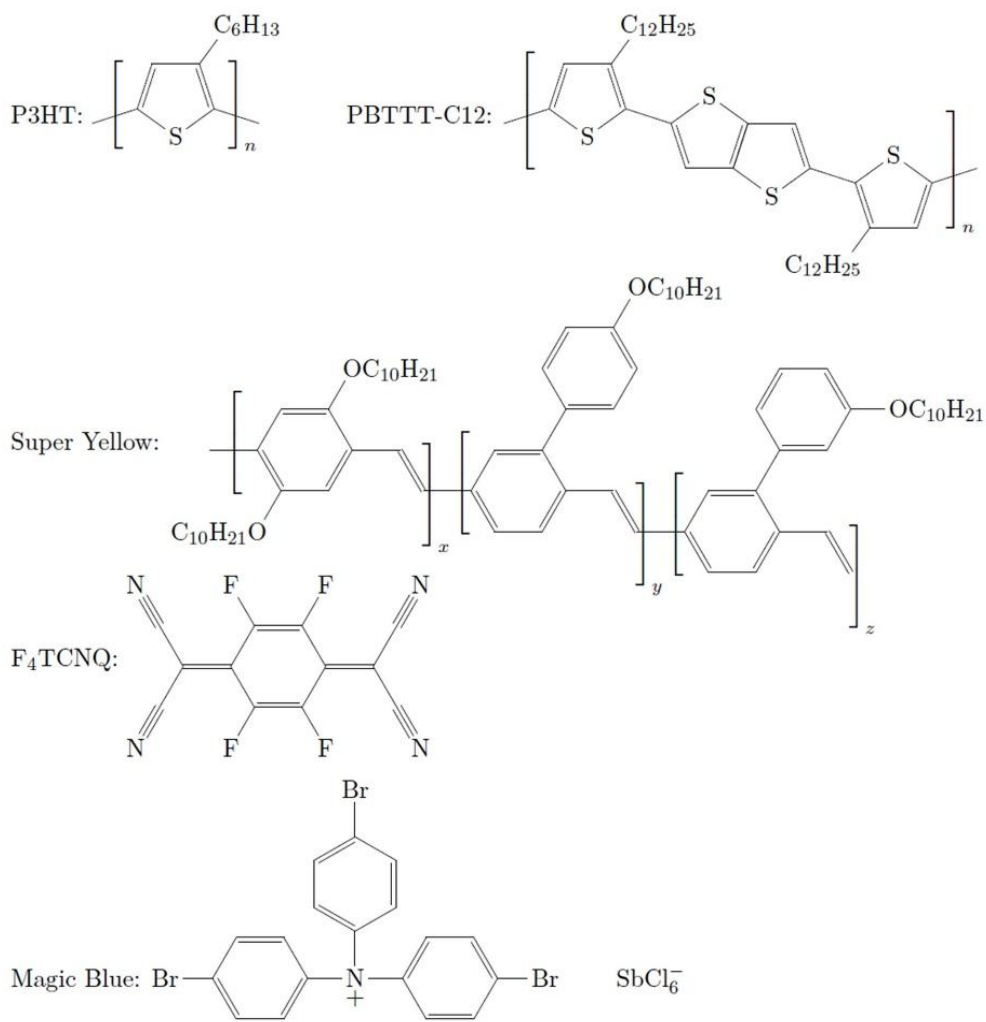

Figure S1. Structures of the materials used in this work.

**Film fabrication:** For thin film deposition, 35 μL of the prepared solution was statically spin-coated onto interdigitated gold electrodes (ED-IDE3-Au from MicruX Technologies) with a 5 μm inter-electrode spacing. The spin-coating protocol consisted of two steps: 1000 rpm for 60 s followed by 3000 rpm for 30 s.

**Doping process:** Here we have used two different doping methods. In the sequential method, a solution of dopant is dynamically spincoated (3000 rpm for 60 s) on the film. In the bulk method, both the dopant and the polymer were mixed in the same solvent and spincoated statically on the substrates.

**Electrical characterization:** Electrical conductivity was measured using a Keithley 2636B Source Meter Unit inside a high-vacuum cryostat. For temperature-dependent measurements, the temperature was varied in 2 K increments between 80 K and 200 K, allowing enough time at each step to ensure thermal equilibrium. Both cooling and heating cycles were measured to verify the absence of hysteresis in conductivity. To ensure good thermal contact and accurate temperature readings, a thin layer of GE varnish (IMI 7031) was applied between the substrate and the cryostat mounting stage.

High-field conductivity measurements were performed using either fast IV sweeps, using a low number of power line cycles (NPLC) per measurement point or pulsed techniques. The choice of method depended on the sample resistance. For medium- to high-resistance samples, fast IV sweeps using an NPLC of 0.001 were sufficient to avoid Joule heating. For low-resistance samples, significant Joule heating occurred at high fields. In these cases, we employed a pulse technique to mitigate thermal effects, as shown in Figure S2a.

In the pulsed method, a series of 20 ms voltage pulses were applied with 2 s intervals between pulses to allow for thermal relaxation. During each pulse, 100 data points were recorded with a 0.2 ms interval. As shown in Figure S2b, the conductance increases over the course of a pulse due to heating.

To estimate the Joule heating-free conductivity, we extrapolated the conductance to time t=0, using a model based on transient heat diffusion. Assuming pulse durations $\tau_p < d^2/D$ (where $d$ indicates the thickness of the glass substrate and $D$ the thermal diffusion constant of glass), the temperature can be estimated using the parameter $\eta = \left(A\sqrt{\pi\kappa\rho C_p}\right)^{-1}$, where A is the area of the device, κ is the heat conductivity of glass, $Cp$ is the heat capacitance of glass and ρ the density of glass. The temperature in the glass can be approximated from the one-dimensional diffusion equation, where at the device side of the glass (z = 0), ∂T/∂z = 0, so that heat can only be transported through the glass and not through the vacuum above. Furthermore, for a short pulse, the thickness of the glass can be treated as infinite. The solution for the temperature-profile is a Gaussian, such that at z= 0:

$$T(t) = T_0 + \eta \int_0^t \frac{P(t')}{\sqrt{t'}} dt' \qquad (1)$$

Where $T_0$ is the temperature before the pulse and P(t) is the power as a function of time measured from the start of the pulse. For glass at room temperature the following constants: κ = 1 $Wm^{-1}K^{-1}$; $C_p$ = 880 $J\,Kg^{-1}K^{-1}$; ρ = 2.4 × $10^3 kgm^{-3}$ result in η =55.5 $K\,W^{-1}s^{-0.5}$ for a 7 $mm^2$ sample at 300 K. Now the temperature dependence of the conductivity is estimated by a Taylor expansion up to first order around the initial temperature since no large heating at short pulses is expected:

$$\sigma(F_0, T) \approx \sigma(F_0, T_0) + \frac{\partial \sigma}{\partial T}(F_0, T_0) \times (T(t) - T_0)$$

$$\approx \sigma(F_0, T_0)\left(1 + (T(t) - T_0)\right)\frac{\partial}{\partial T}\log\sigma_{ohm} \qquad (2)$$

The resulting curve has two fitting parameters, first the initial conductivity $\sigma(F_0, T_0)$ in which we are interested, and secondly the heating parameter $\eta$ which is calculated above.

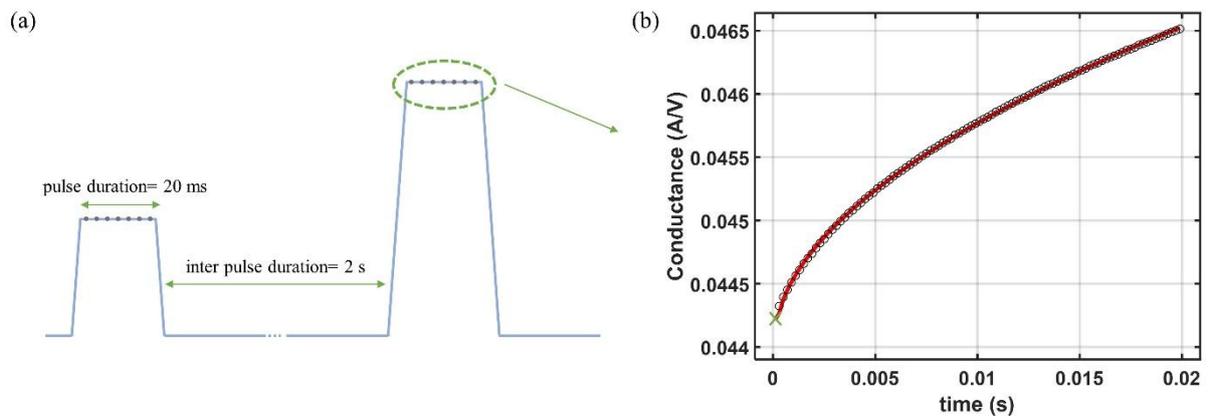

**Figure S2.** Pulse measurement technique and the heat correction for joule heating at high electric fields.

## 2. Contact resistance

Non-ohmic contacts between electrodes and the organic semiconductor can contribute significantly to the total device resistance, particularly at low to intermediate electric fields. In this regime, the apparent field dependence of conductivity may be influenced or even completely caused by a field-dependent contact resistance, if present. To assess the impact of contact resistance in our system, we performed in-operando Kelvin Probe Force Microscopy (KPFM) measurements using an Oxford Instruments Jupiter XR atomic force microscope (AFM) operated in heterodyne mode. The measurements were conducted on a representative sample of regioregular P3HT, sequentially doped with $F_4$TCNQ at a concentration of 0.02 mg/mL, under various applied electric fields. The measurements were performed on interdigitated electrodes with the fast sweep direction along the field, averaging over all lines along the electrode direction to generate an accurate potential profile between the differently poled electrodes.

Figure S3a and S3b display the surface potential and corresponding height profiles at applied bias of 7 V. At low bias, we observe a step-like drop in the potential profile near the central electrode, indicative of contact resistance whereas the outer contacts exhibit a less pronounced potential drop, suggesting, somewhat surprisingly, the presence of a hole extraction barrier but not of a hole injection barrier. To investigate the field dependence of the contact resistance, we quantified the ratio of the potential step to the total channel potential drop at varying electric fields. As shown in Figure S3c, the ratio of contact to total resistance remains largely constant, and only slightly drops towards higher fields. While these results do not provide a definitive measure of contact resistance in our experiments, we conservatively estimate that it contributes maximally 20% of the total resistance in our devices. Accordingly, in our analysis, we focus on higher electric fields for which the observed increase in conductivity significantly exceeds 20%, thus ensuring that the primary contribution is from field-activated transport rather than contact artifacts. For example, as shown in Figure S3d, the conductivity increases by more than 20% at fields as low as $10^5$ V/m, a value far below the field range where strong nonlinearities are observed. This confirms that the pronounced high-field dependence of conductivity in our samples, of up to several orders of magnitude, cannot be attributed to contact resistance.

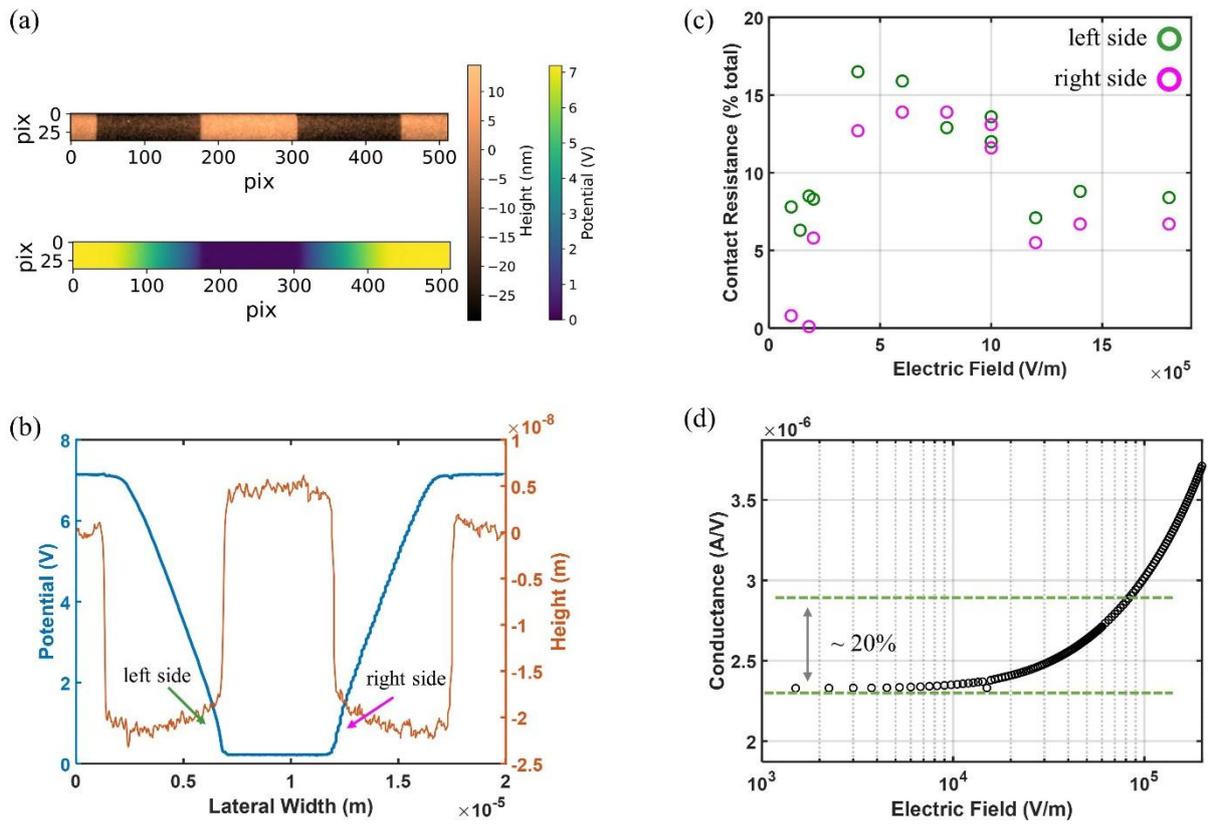

**Figure S3.** (a) and (b) Height and potential profile at the bias of 7 V. (c) the ratio of the potential drop as a function of bias voltage/electric field. (d) conductance as a function of electric field.

## 3. Heat balance model

While the concept of the effective temperature proposed by Marianer and Shklovskii is well established for disordered systems, there has been no physical justification for the resulting expression describing the effective temperature:

$$T_{eff}^{\beta} = T_{latt}^{\beta} + \left(\frac{\gamma q \alpha F}{k_B}\right)^{\beta} \tag{3}$$

A physically transparent way to rationalize the concept of effective temperature is to view it as the balance between carrier heating by the applied field and cooling through energy transfer to the lattice. The power input per unit volume from Joule heating is:

$$\dot{Q}_H = \sigma F^2 = q n \mu F^2 \tag{4}$$

where $\sigma = qn\mu$ is the conductivity, $\mu$ is the mobility and $n$ is the charge carrier density. The competing cooling flux can be approximated as the relaxation of the carrier distribution towards the lattice temperature,

$$\dot{Q}_C = n k_B \frac{T_{eff} - T_{latt}}{\tau} = q n \mu F^2 \tag{5}$$

where $\tau$ is the thermal relaxation time. Equating the two contributions gives the heat-balance relation

$$q n \mu F^2 = n k_B \frac{T_{eff} - T_{latt}}{\tau} \tag{6}$$

To evaluate $\tau$, we assume hopping transport, where each phonon-assisted hop acts as a thermalization event. The hopping diffusion coefficient is then expressed as $D = \alpha^2/6\tau$ where $\alpha$ is the typical hop length. Using the Einstein-Smoluchovski relation $D = k_B T_{eff} \mu / q$ we obtain:

$$\frac{1}{\tau} = \frac{6 k_B T_{eff} \mu}{q \alpha^2} \tag{7}$$

By substituting Eq. 6 in Eq. 7 we obtain:

$$q n \mu F^2 = \frac{6 k_B^2 T_{eff} n \mu}{q \alpha^2} (T_{eff} - T_{latt}) \tag{8}$$

and with further simplification we obtain the final expression:

$$T_{eff} = \frac{T_{latt} + \left(T_{latt}^2 + \left(\gamma \frac{q \alpha F}{k_B}\right)^2\right)^{0.5}}{2} \tag{9}$$

where $\gamma \approx 0.8$ absorbs numerical prefactors of order unity. This formulation provides a physically motivated foundation for the effective temperature concept, showing explicitly how electric field strength and localization length control the deviation of carrier temperature from the lattice.

## 4. Raw experimental data

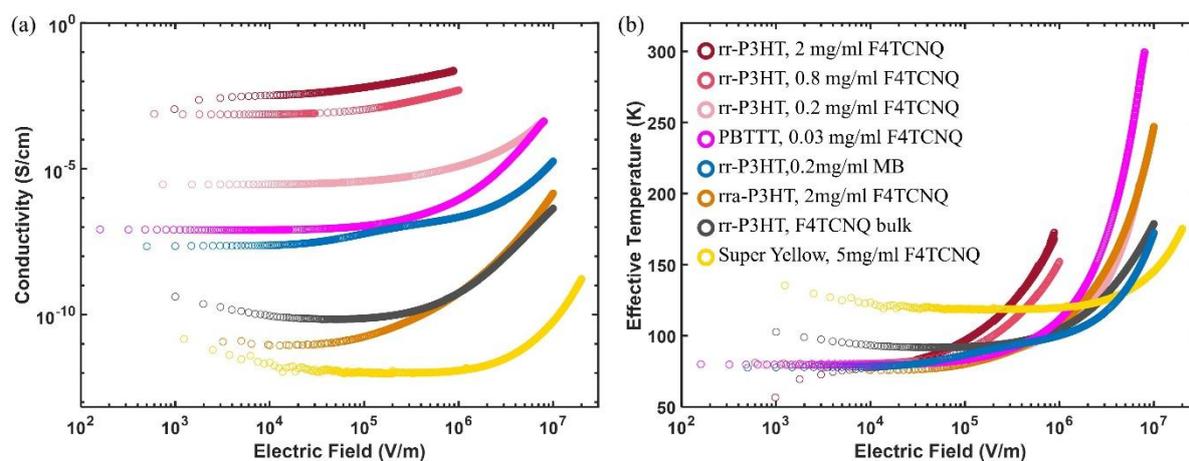

Figure S4. Field dependence of (a) conductivity and (b) effective temperature in all material systems investigated in this work.

## 5. Tight binding model details

The basis orbitals are assumed ellipsoids of the form

$$\psi(x,y,z) = \frac{1}{\sqrt{\pi l_x l_y l_z}} \exp\left(-\sqrt{\left(\frac{x}{l_x}\right)^2 + \left(\frac{y}{l_y}\right)^2 + \left(\frac{z}{l_z}\right)^2}\right)$$

with $(l_x, l_y, l_z) = (0.88, 0.88, 1.3) \cdot 0.1$ nm to account for the enhanced overlap in the π-π-stacking direction. At this point it should be mentioned that these values are consistent with the offset value $\lambda_S = 0.2$ nm (orbital localization) since the localization length provided by the IPR reflects a side length of a box, whereas the values $l_x, l_y$ and $l_z$ are radii (i.e. half diameters). It therefore holds approximately $2\sqrt[3]{l_x \cdot l_y \cdot l_z} \approx \lambda_S$. These orbital wavefunctions are rotated according to the monomer (=site) orientations. The transfer integrals were calculated by overlaps $S_{ij}$ of two wavefunctions scaled by a prefactor $\hbar v_0$ with $v_0 = 2 \cdot 10^{14} \frac{1}{s}$, corresponding to $\hbar v_0 \cong 125$ meV; we use a relatively high value for $v_0$ to obtain a reasonable upper limit for the delocalization and hence the field dependence of the wavefunctions. Intra-chain transfer integrals of neighboring monomers that are connected in a straight conformation are enhanced by a factor of $f_{ich} = 1.3$. The on-site energies for the amorphous phase were taken to be 80 meV and 45 meV for the aggregated phase.

The (simplified) entangled polymer morphology was generated by nucleating 10% of the lattice sites and from thereon grow polymer chains with maximum length drawn from a Flory-Schultz distribution with number averaged number of repeat units per chain of 200 or until they hit a dead end. The growth is done monomer by monomer with an orientation that is randomly chosen, each weighted by the resulting summed interaction energy with the surrounding monomers as

$$p \propto \exp\left(-\frac{E_{int}}{k_B T}\right) \tag{9}$$

The energetic penalties (values > 0) / gains (values < 0) for intrachain orientations are $(E_{straight}, E_{twist}, E_{bend}) = (0, 4, 1)$ (arbitrary units). The interchain conformation energies between two neighboring monomers are chosen as $(E_{face-face}, E_{face-edge}, E_{edge-edge,\parallel}, E_{edge-edge,\perp}) = (-8, 8, -3, 0)$. The monomers are aligned in z-direction by an alignment field of $(F_x, F_y, F_z) = (0,0,2)$ and the monomer directions by an alignment field into x and y-direction by $(F_x, F_y, F_z) = (2,2,0)$. The probability of the nucleation of new chains are length-weighted and an over-filling factor of 7 was taken to avoid larger voids within the final morphology. These parameters lead to small but homogeneously distributed aggregates (Figure S5) of sizes up to approximately 4 nm. The average aggregate size $r_c$ is calculated as average over the typical aggregation size of a single chain calculated as $r_{typ} = \sum_i |r_i - r_{COM}|/\sum_i 1$ with the sum running over all monomers with position $r_i$ of a single chain and the chains center of mass $r_{COM}$. Due to the finite box and large overfilling factor, the average value $r_c$ is however low due to overrepresentation of small chains (Figure S6). This is also reflected by the resulting chain length distribution in Fig S5, which deviates from the idealized input Flory-Schulz distribution and gives a lower number averaged molecular weight of $M_n = 26.3$. For more details on the model and its parameters, see [https://doi.org/10.1038/s41563-025-02207-9].

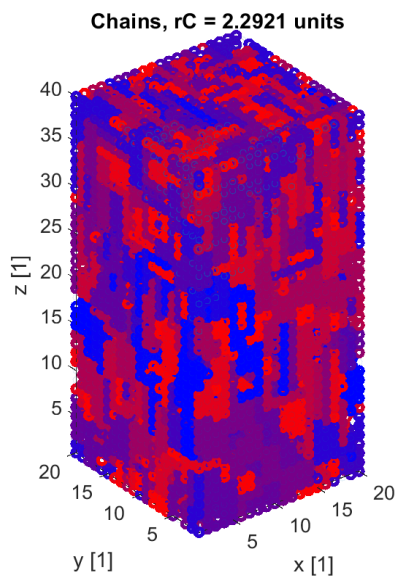

Figure S5. Generated aggregated morphology. Chains in blue are longer, chains in red are shorter

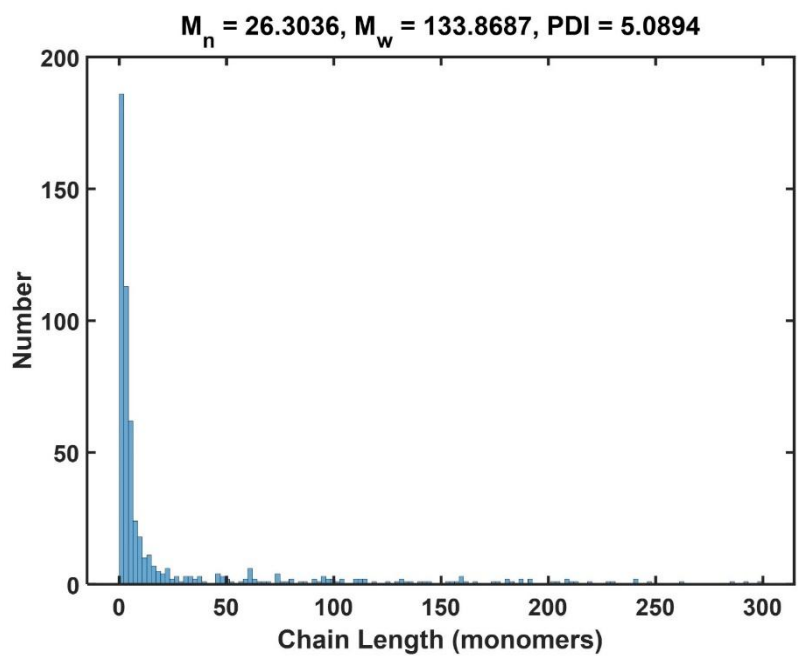

Figure S6. Chain length distribution